\documentclass[final,3p,times,twocolumn,number,sort&compress]{elsarticle}

\graphicspath{{../Figures-EPS-Numbers/}{.}}
\usepackage[]{color}
\usepackage[]{graphicx}
\usepackage[]{enumerate}

\newcommand{\comment}[1]{}

\journal{Nuclear Instruments and Methods in Physics Research A}

\begin{document}

\begin{frontmatter}

\title{Silicon photomultiplier readout of a monolithic 270$\times$5$\times$5\,cm$^3$ plastic scintillator bar for time of flight applications}

\author[tu]{Tobias P. Reinhardt}
\author[hzdr,tu]{Stefan Gohl}
\author[hzdr,tu]{Stefan Reinicke}
\author[hzdr]{Daniel Bemmerer}
	\ead{d.bemmerer@hzdr.de}%
\author[hzdr,tu]{Thomas E. Cowan}
\author[hzdr]{Klaus Heidel}
\author[tu,hzdr]{Marko R\"oder}
\author[hzdr]{Daniel Stach}
\author[hzdr]{Andreas Wagner}
\author[hzdr]{David Weinberger}
\author[tu]{Kai Zuber}
\author[]{for the {{R$^3$B}} collaboration}
\address[tu]{Technische Universit\"at Dresden, Institut f\"ur Kern- und Teilchenphysik, Zellescher Weg 19, 01062 Dresden, Germany}
\address[hzdr]{Helmholtz-Zentrum Dresden-Rossendorf (HZDR), Bautzner Landstr. 400, 01328 Dresden, Germany}

\begin{abstract}
The detection of 200-1000\,MeV neutrons requires large amounts, $\sim$100\,cm, of detector material because of the long nuclear interaction length of these particles. In the example of the NeuLAND neutron time-of-flight detector at FAIR, this is accomplished by using 3000 monolithic  scintillator bars of 270$\times$5$\times$5\,cm$^3$ size made of a fast plastic. Each bar is read out on the two long ends, and the needed time resolution of $\sigma_t$ $<$ 150\,ps is reached with fast timing photomultipliers. In the present work, it is investigated whether silicon photomultiplier (SiPM) photosensors can be used instead. Experiments with a picosecond laser system were conducted to determine the timing response of the assembly made up of SiPM and preamplifier. The response of the full system including also the scintillator was studied using 30\,MeV single electrons provided by the ELBE superconducting electron linac. The ELBE data were matched by a simple Monte Carlo simulation, and they were found to obey an inverse-square-root scaling law. In the electron beam tests, a time resolution of $\sigma_t$ = 136\,ps was reached with a pure SiPM readout, well within the design parameters for NeuLAND. 
\end{abstract}

\begin{keyword}
Silicon Photomultiplier \sep Plastic scintillator \sep time resolution \sep FAIR \sep ELBE \sep neutron detection \sep picosecond laser system \sep time-of-flight detector
\end{keyword}

\date{\today}

\end{frontmatter}

\section{Introduction}

Nuclear reactions involving nuclei close to or beyond the neutron drip line are relevant for the synthesis of the heavy chemical elements \cite{Wiescher12-ARAA}. One of the experimental methods to investigate such reactions is the invariant mass method, which requires a kinematically complete measurement. By detecting
and identifying the products of the nuclear reaction in question and determining their momenta, the invariant mass of the system is reconstructed. Due to an abundance of  neutrons in the typical nuclei under investigation, this technique usually involves
neutron detection. 

The R$^{3}$B (Reactions with Relativistic Radioactive \linebreak Beams) collaboration aims to study the reactions of such exotic nuclei \cite{Aumann07-PPNP}. At the present R$^{3}$B setup in cave C of GSI Helmholtz Centre for Heavy Ion Research, the Large Area Neutron Detector (LAND)~\cite{Blaich92-NIMA} is used. This device covers 2$\times$2\,m$^2$ area and has reached a single-neutron efficiency of 90\% for 0.5\,GeV neutrons, with a typical time-of-flight resolution of $\sigma$ = 250\,ps \cite{Boretzky03-PRC}. Similar but smaller detectors for high-energy neutrons exist at radioactive ion beam facilities in the United States \cite{Baumann05-NIMA}, China \cite{Yu09-CPC},  Japan \cite{Kobayashi13-NIMB}, and at the COSY-TOF spectrometer in Jülich, Germany \cite{Karsch01-NIMA}.
The use of the LAND detector has enabled a wide physics program, for light \cite{Caesar13-PRC,Roeder16-PRC,Jonson04-PhysRep} as well as heavy nuclei \cite[and references therein]{Aumann98-ARNPS,Aumann05-EPJA}.%

The ongoing construction of a new infrastructure for producing
radioactive ion beams named FAIR (Facility for Antiproton and Ion Research) in Darmstadt will provide for very exotic beams. Using the future R$^3$B setup at FAIR, several important topics will be addressed \cite{Aumann07-PPNP,Reifarth13-NPA6}. These include the dipole response and giant resonance studies of very neutron rich heavy nuclei, the precise determination of the strength function at the particle threshold for nuclei of astrophysical relevance,  nuclear structure investigations via the quasifree scattering method, and the investigation of unbound resonances for lighter particles. This new setup requires a more powerful neutron detector for the future R$^3$B setup, which is called NeuLAND \cite{NeuLAND-TDR11}. 

The properties aimed for with NeuLAND are unprecedented for a fast neutron (0.2-1.0~GeV) array: at least 90\% detection efficiency for single neutrons at 0.2~GeV energy, 
a very large angular coverage of 80~mrad at a distance of 15.5\,m to the reaction target,
an excellent time resolution of $\sigma_t$ = 150\,ps and
the needed granularity to achieve good invariant mass resolution, e.g.,
$\sigma_E$ = 20 keV at 100 keV excitation energy above the threshold for medium mass systems.
Also, the setup shall be able to identify multi-neutron
events with up to four neutrons per event, and correctly reconstruct their momenta.

\begin{figure*}[tb]
\includegraphics[angle=0,width=\textwidth]{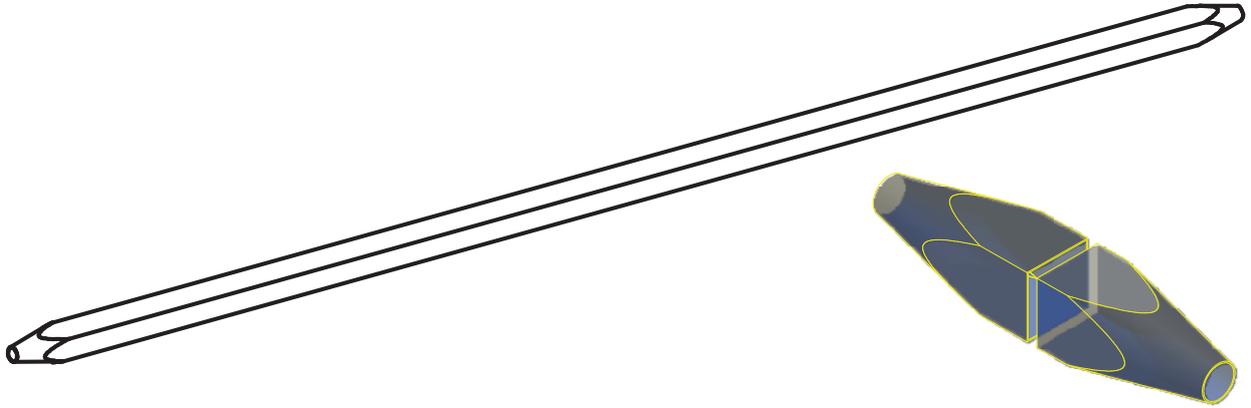}
\caption{\label{fig:NeuLAND} NeuLAND bar. The left side shows  the entire, 270\,cm long NeuLAND bar. The right side shows the two tapered sides converting from 5$\times$5\,cm$^2$ square shape to $d$ = 2.5\,cm circular shape.}
\end{figure*}

After deeply studying a detection approach based on multigap resistive plate chambers \cite{Yakorev11-NIMA,Roeder12-JINST,Elekes13-NIMA,Roeder14-EPJA}, the R$^3$B collaboration has decided to use a plastic scintillator-based detection concept instead, based on the better multihit reconstruction in such a fully active detector \cite{NeuLAND-TDR11}. However, this approach poses a challenge due to the need to procure, and maintain, not only large amounts of fast plastic scintillator, but also 6000 fast timing photomultiplier tubes (PMTs). 

The present work investigates an alternative approach, using silicon-based photosensors, so-called silicon photomultipliers (SiPMs \cite{Buzhan03-NIMA}) instead of fast timing PMTs to read out the large NeuLAND scintillator bars. It is already known that SiPMs may provide competitive time resolutions for the case of small scintillators \cite{Stoykov12-NIMA}, and also that large scintillator bars may be read out by SiPMs \cite{Balagura06-NIMA}. Here, it is investigated under which conditions SiPM readout of a large scintillator is possible with competitive time resolution. Similar efforts by other groups have been reported elsewhere \cite{Kaplin15-PhysProc}.

This work is organized as follows. The detector assembly studied, consisting of a NeuLAND plastic scintillator bar, the SiPM, and the preamplifier, is described in sec.\,\ref{sec:Detector}. Section\,\ref{sec:PiLas} discusses tests of the SiPM with a picosecond laser system. Experiments at the superconducting electron linac ELBE, testing a NeuLAND bar read out on both ends by SiPMs, are presented in sec.\,\ref{sec:ELBE}. These data are interpreted with the aid of a Monte Carlo simulation in sec.\,\ref{sec:MonteCarlo}. Section \ref{sec:Arrays} describes ELBE experiments with small arrays of four 6$\times$6\,mm$^2$ SiPMs. A summary and outlook are given in sec.\,\ref{sec:Summary}.

\section{Detector}
\label{sec:Detector}

The precise geometry and chemical composition of the NeuLAND plastic scintillator bars has been defined in the NeuLAND technical design report \cite{NeuLAND-TDR11}, and a number of units have already been acquired and installed. The present challenge, therefore, is to develop a SiPM-based readout scheme satisfying ambitious efficiency and time resolution goals without any modification of the scintillator and light-guide.

\subsection{NeuLAND plastic scintillator bar}
\label{subsec:NeuLAND}

The main part of each NeuLAND bar (fig.\,\ref{fig:NeuLAND}) is 250\,cm long and has a square area of 5$\times$5\,cm$^2$. On each side, an additional 10\,cm long section connects the square 5$\times$5\,cm$^2$ area to a circular area of 2.5\,cm diameter, designed to be instrumented by a 1'' diameter PMT. This brings the total length of the NeuLAND bar to 270\,cm, and the area to be instrumented on each side is $25^2\pi/4$ = 491\,mm$^2$. 

The scintillation material is the fast plastic polyvinyltoluene with the trade name RP408, equivalent to BC408 and EJ200. According to the data sheet, it has a scintillation rise time $\tau_{\rm rise}$ = 0.9\,ns, a decay time of $\tau_{\rm decay}$ = 2.1\,ns, an emission spectrum peaked at 425\,nm wavelength, a refractive index of $n$ = 1.58, and a scintillation efficiency of 10,000 photons/MeVee (MeV electron equivalent). This material has been selected for its high hydrogen content of $n_{\rm H}=5.17\times10^{22}\,{\rm cm}^{-3}$ favoring neutron detection, its good timing capabilities and mechanical properties. 
 
In the default NeuLAND configuration, each bar is instrumented with two fast timing PMTs of 1'' diameter. The average of the two PMTs provides the "stop" signal for the time of flight determination of relativistic neutrons in the R$^3$B experiment.  The lateral size of 5$\times$5\,cm$^2$ is a compromise between the needed granularity in the y and z axes, on the one hand, and economical considerations to limit the number of photosensors and readout channels, on the other hand. The time difference between the two PMTs on the two ends gives some position sensitivity along the long side of the NeuLAND bar that matches the 5\,cm granularity of the bars.

\subsection{Silicon photomultiplier photosensors used}
\label{subsec:SiPMList}

For the tests, a number of silicon photomultipliers of various manufacturers were used, ranging in size from 1$\times$1 - 6$\times$6 mm$^2$ and covering several series (Table~\ref{Table:SiPM}). In addition, in order to increase areal coverage, small arrays of four 6$\times$6 mm$^2$ SiPMs each were built and studied.

\begin{table}[tb]
\begin{tabular}{|l|c|r|r|}
\hline
\bf Producer  & \boldmath $A$  & \bf Pitch & \bf\boldmath $U_{\rm BD}$ \\ 
\bf and type & \bf\boldmath [mm$^2$] & \bf\boldmath [$\mu$m] & \bf [V] \\ \hline
Ketek PM1150 &  1$\times$1 & 50 & 25 \\ 
Ketek PM3350 & 3$\times$3 & 50 & 27 \\ \hline
Excelitas C30742-33 & 3$\times$3 & 50 & 98 \\ \hline
SensL C-series & 6$\times$6 & 35 & 25 \\ \hline
FBK NUV & 6$\times$6 & 40 & 33 \\
\hline
\end{tabular}
\caption{\label{Table:SiPM} List of SiPM photosensors used for the test, their active area $A$, pitch size, and typical breakdown voltage $U_{\rm BD}$.}
\end{table}
\subsection{In-house developed preamplifier}
\label{subsec:Preamp}

\begin{figure}[tb]
\begin{center}
\includegraphics[angle=0,width=\columnwidth]{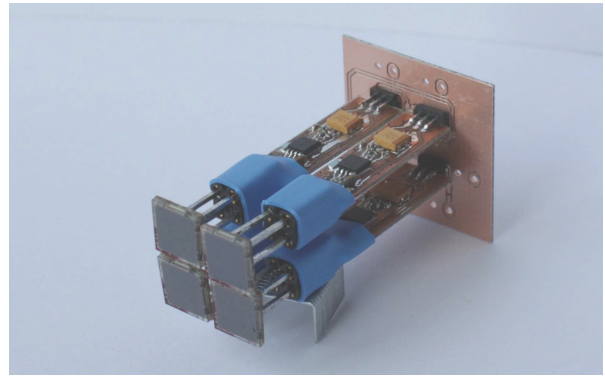}
\end{center}
\caption{\label{fig:Preamplifier} Photograph of the preamplifier board, complete with four 6$\times$6\,mm$^2$ SiPMs. When in use, the SiPMs are separated from the board by a neoprene layer that is penetrated by the pin connectors of the SiPMs.}
\end{figure}

Due to the limited availability of specially adapted commercial preamplifiers for SiPMs, a new preamplifier was developed in-house at HZDR (fig.\,\ref{fig:Preamplifier}). The preamplifier is capacitively coupled to the SiPM anode. The coupling capacitance of typically 100\,nF is high enough to amplify also low frequency parts of the detector signal, which is done in order to avoid a bipolar output signal. 

The preamplifier input stage is implemented as a standard grounded-emitter circuit for voltage and current amplification in both the low and high frequency domains. The operating point is set by a base voltage divider. This stage has a constant impedance of 50\,$\Omega$ over a very wide frequency range of 10$^3$-3$\times$10$^9$\,Hz. The constant input stage impedance insulates the preamplifier from SiPM capacitance changes when cells break through. The high impedance at the output of the first-stage grounded-emitter circuit is addressed by the second amplification stage. It is implemented as a grounded-collector circuit, which can directly serve a 50\,$\Omega$ impedance load. The anode preamplifier has a bandwidth of 3\,GHz, enabling rise times in the range from 0.1\,ns (0\,dB gain) to 1.1\,ns (30\,dB gain). 

For the case of the SensL C-series SiPMs, in addition to the "energy" output at the SiPM anode a second output is provided for fast timing purposes. This output signal is much smaller than the anode signal. Therefore, it is amplified by two current feedback operational amplifiers of type THS3202 that are employed as inverting amplifiers. Each of the two stages for the fast output has a gain of a factor of 10. The THS3202 is characterized by a low input impedance (11\,$\Omega$ in inverting mode) and a high amplification bandwidth (2\,GHz for unity gain). 

This preamplifier is implemented for each SiPM unit separately. For the case of the small SiPM arrays described below, the preamplifier outputs from the individual SiPMs are added by a simple summing amplifier. 

In addition to the preamplification, the in-house developed preamplification scheme provides two low-bandpass filters for the SiPM bias voltage: one on the preamplifier board, the second on the summing board, if applicable. 

Each preamplifier board serving one SiPM and also each summing board include their own linear regulator for the electronics power supply (typically 6\,V).

\section{Picosecond laser measurements}
\label{sec:PiLas}

\begin{figure}[tb]
\includegraphics[angle=0,width=\columnwidth]{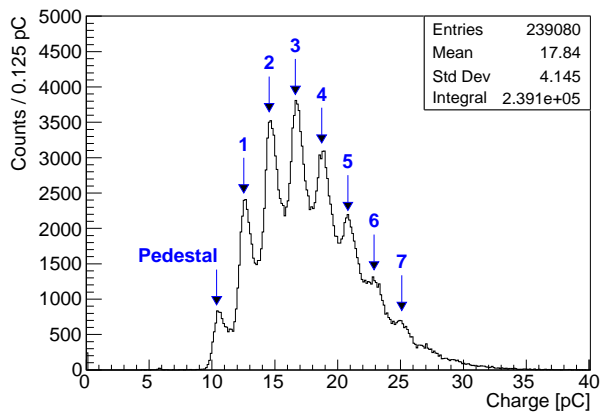}
\caption{\label{fig:PiLas_ChargeSpec} Charge spectrum taken with a 6$\times$6\,mm$^2$ SiPM prototype (SensL) and the PiLas laser at $U_{\rm OV}$ = 2.0\,V. The laser light has been attenuated far enough to make single fired cells (from one to seven cells) visible. }
\end{figure}

Each SiPM, in turn, was illuminated with a picosecond laser system, hereafter called PiLas \cite{Gohl14-Diplom}. A PiL042XSM unit (Advanced Laser Diode Systems, Berlin, Germany) was used.  The PiLas has a wavelength of 420.8 nm (spectral width 1.8 nm). According to the data sheet, the timing of the pulse is defined to 38\,ps (FWHM) with respect to the trigger output of the PiLAS controller. 

The light output from the PiLAS laser head was coupled by way of an optical fiber to a light tight box. Inside the box there was an exchangeable optical glass filter, supplying an optical attenuation factor varying between 30-100. In addition to the optical filter, the intensity could also be varied by changing the "tune" parameter on the PiLas, and a dynamic range of 1-500 fired pixels was reached for the assembly consisting of PiLas and optical filter. Subsequently, the SiPM and its preamplifier (separated from each other by a light-tight neoprene layer) were attached. This setup prevented parasitic light from reaching the SiPM. 

As a first step, oscilloscope-based measurements were performed. To this end, the trigger output of the PiLAS controller was connected to one channel of a 12-bit 1\,GHz, 2.5\,GS/s digital oscilloscope (LeCroy HDO6104MS); the oscilloscope was set to trigger on this channel. The SiPM preamplifier output was connected to a linear fan-out unit (Philips Scientific PS 748), then to another channel of the oscilloscope. 

The waveforms were recorded and analyzed offline, using several different software emulations of a constant fraction discriminator. The best timing performance was found with a linear interpolation of the leading edge of the signal, using a fraction of 22\% of the full signal height. Slightly worse performance was obtained by fitting either the entire leading edge or large parts of it with an arcustangens function.

Once the proper timing fraction and charge integration window were set based on the oscilloscope measurements, data were taken with a second and third branch of the data acquisition system after the fan-out unit: In the second branch, the timing was determined with an in-house made constant fraction discriminator (CFD, model HZDR CFT 96687) which allowed to set a threshold as low as 7\,mV. For the present purposes, a  threshold of 15\,mV was used. The CFD output was fed into a 25\,ps multihit time-to-digital converter (TDC, model CAEN V1290). 

The third branch after the fan-out unit consisted of a delay and then a 25\,fC charge-to-digital converter (QDC, CAEN V965) to determine the signal charge. The trigger for the entire data acquisition system was built inside a field programmable gate array (FPGA, model CAEN V1495) from a logical AND of the SiPM signal and the laser trigger, with the latter determining the timing. The FPGA also supplied the gate signal (length 60\,ns) for the QDC. 

The data acquisition was controlled and the data were recorded to hard disk by a GSI multi-branch system \cite{Essel00-IEEE}. The data stream was monitored by the GO4 online analysis system  \cite{Adamczewski11-IEEE}. The same tool was used to convert the list mode data files to ROOT files for further analysis.

As a first step, the PiLas intensity was attenuated by a factor 30-100, so that only one or a few pixels fired in the SiPM under study (fig.\,\ref{fig:PiLas_ChargeSpec}). The operating voltage $U$ was then varied in 0.1-1.0\,V steps, starting from the datasheet recommended value. For each voltage setting, the charge spectrum was used to experimentally determine the pedestal given by the sum of the dark current of the SiPM, on the one hand, and an offset current added by the QDC, on the other hand, as well as the QDC calibration in pC per fired pixel. 

Both the offset (pedestal) and the slope of the calibration were found to depend significantly on $U$. This allowed to experimentally re-determine the breakdown voltage $U_{\rm BD}$. To this end, the charge of the n-th fired pixel was plotted as a function of $U$, and $U_{\rm BD}$ was determined as the voltage where the several lines cross \cite{Vinke09-NIMA,Gohl14-Diplom}. The measured breakdown voltages deviated by less than 0.1\,V from the data sheet values for the relatively modern SiPMs discussed here, and the data sheet numbers were used henceforth. For the purpose of the further discussion, instead of the operating voltage $U$ the so-called overvoltage $U_{\rm OV}$ given by $U_{\rm OV} = U-U_{\rm BD}$ is used.

\begin{figure}[tb]
\includegraphics[angle=0,width=\columnwidth]{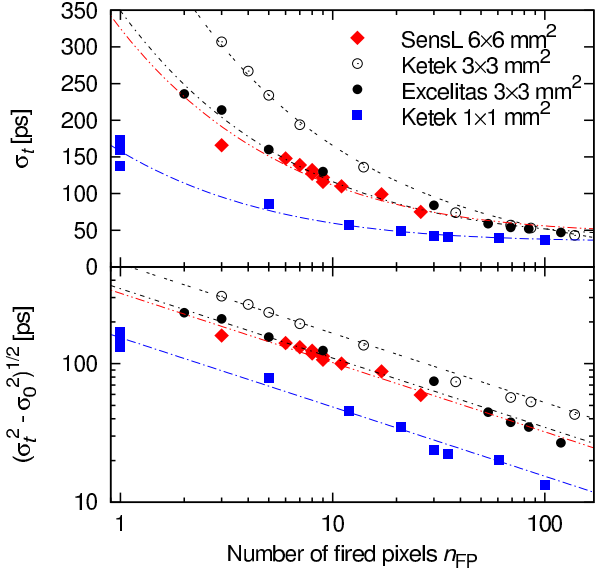}
\caption{\label{fig:PiLas_FiredPixels}  Top panel: Time resolution $\sigma_t$ measured with the PiLas system, as a function of the number of fired pixels $n_{\rm FP}$, for SiPM prototypes with different active areas. Bottom panel: Time resolution after quadratic subtraction of the fitted offset $\sigma_0$.  The lines in both panels are fits using eq.~(\ref{eq:sigma_t}). See text for details.}
\end{figure}

As a next step, for several of the SiPMs studied here, the time resolution of the preamplified signal as a function of the number of fired pixels $n_{\rm FP}$ was determined with reference to the PiLas trigger output (fig.\,\ref{fig:PiLas_FiredPixels}, top panel). It should be noted that for this test, all emitted photons are strictly correlated within the PiLas time resolution, which is a different scenario from what is true when detecting scintillation light, see sec.~\ref{sec:MonteCarlo} below. For the small 1$\times$1\,mm$^2$ SiPM, already for  $n_{\rm FP}$ = 1-2, the NeuLAND aim of  $\sigma_t \leq$ 150\,ps is reached. When considering all SiPMs studied here, $\sigma_t \leq$ 150\,ps is found for $n_{\rm FP} \geq$ 15. For large numbers of firing pixels  $n_{\rm FP} \geq$ 100, the time resolution reaches its optimum at $\sigma_t \approx 40$\,ps (fig.\,\ref{fig:PiLas_FiredPixels}, top panel). 

In order to experimentally verify the previously reported \cite{Collazuol07-NIMA,Ronzhin10-NIMA,Acerbi14-IEEE} antiproportionality $\sigma_t \propto 1/\sqrt{n_{\rm FP}}$, the data are fitted with the following modified function that includes a possible offset $\sigma_{0}$ for the time resolution: 
\begin{equation}\label{eq:sigma_t}
\sigma_t = \sqrt{\sigma_{0}^2 + \frac{\sigma_1^2}{n_{\rm FP}}}
\end{equation}
Here, the constant $\sigma_1$ represents the fitted time resolution for just one fired pixel. $\sigma_1$ depends on the size of the SiPM, thus on the number of pixels included in the device, and on the dark count level. 

The offset $\sigma_{0}$ characterizes the jitter of the system and should be of the same order of magnitude as the time resolution of the PiLas laser. The fitted values for $\sigma_0$ for three of the four prototypes shown in fig.\,\ref{fig:PiLas_FiredPixels} are 35-46\,ps, respectively. This confirms the modified inverse square root dependence on $n_{\rm FP}$ given by eq.~(\ref{eq:sigma_t}), for SiPM sizes between 1$\times$1 and 6$\times$6\,mm$^2$, corresponding to a variation of the number of connected SiPM pixels by a factor of 30.

A fourth prototype (Ketek 3$\times$3\,mm$^2$) is somewhat noisier, possibly due to imperfect matching to the present preamplifier. In that case, the fit only converges when $\sigma_{0}$ is set to zero, corresponding to the original scaling relation $\sigma_t \propto 1/\sqrt{n_{\rm FP}}$ reported in the literature.  

For the practical purpose of a working NeuLAND photosensor, it can be concluded from the laser measurements that for $\geq$15 correlated firing SiPM pixels, the NeuLAND time resolution aim of $\sigma_t$\,$\le$\,150\,ps is reached for all SiPMs studied. Furthermore, any increase in time resolution due to a larger SiPM area may be compensated by a larger number of fired pixels, i.e. more light. 

\section{Electron beam measurements with a scintillator read out by single SiPMs}
\label{sec:ELBE}

\begin{figure}[tb]
\includegraphics[angle=0,width=\columnwidth]{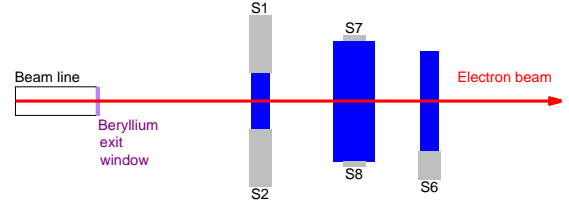}
\caption{\label{fig:ELBE_Setup} Schematic view of the experimental setup at the ELBE electron accelerator.}
\end{figure}

\subsection{Principle of the measurements}

After proving sufficient time resolution for a SiPM irradiated with a picosecond laser (sec.\,\ref{sec:PiLas}), the logical next step is a study of the efficiency and time resolution of a composite system made up of plastic scintillator, SiPM, and preamplifier. Two different plastic scintillator samples were used:
\begin{enumerate}
\item A test bar of 42$\times$11$\times$100\,mm$^3$ EJ-200, selected because of its easy handling and small dimensions.
\item An original NeuLAND bar of 50$\times$50$\times$2700\,mm$^3$ with tapered ends as described in sec.\,\ref{subsec:NeuLAND}.
\end{enumerate}

The two plastic scintillator samples were subsequently exposed to the single-electron beam of the ELBE superconducting electron linac, and the time resolution was measured with reference to the radio-frequency oscillator of ELBE. This technique is well-established \cite{Yakorev11-NIMA,Naumann11-NIMA} and has been previously employed with great success to study 2.0$\times$0.5\,m$^2$ large resistive plate chambers \cite{Yakorev11-NIMA,Roeder12-JINST,Elekes13-NIMA}. 

The 30\,MeV electrons used show an energy loss that is only 15\% higher than that of minimum ionizing particles. The use of single electrons per bunch thus allows a stringent test for the radiation detectors under study, with the detected signal caused by just one, almost minimum ionizing particle. The conclusions of the present test protocol may thus be applied also to more relaxed conditions, e.g. where particles with higher specific ionization or more than one detected particle cause the signal.

\begin{figure}[tb]
\includegraphics[angle=0,width=\columnwidth]{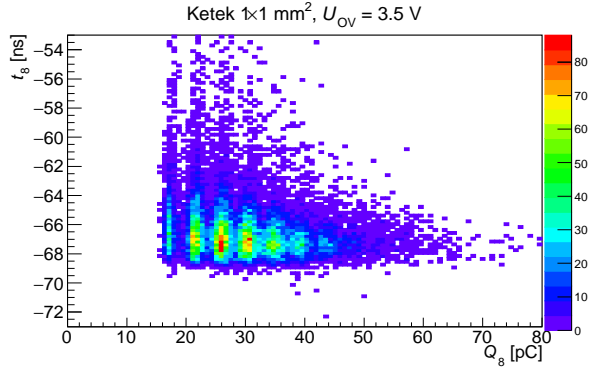}
\caption{\label{fig:Banana}  Time-over-charge plot for a KETEK 1$\times$1\,mm$^2$ SiPM irradiated with the ELBE 30\,MeV electron beam.}
\end{figure}

\subsection{Experimental setup and procedure}

The experimental setup (fig.\,\ref{fig:ELBE_Setup}) is built around the 30\,MeV electron beam from ELBE. The beam intensity was reduced so far that each accelerated electron bunch contained a maximum of one electron, the so-called single-electron mode \cite{Naumann08-Patent,Yakorev11-NIMA,Naumann11-NIMA}. Different from previous applications of this mode of operation at ELBE \cite{Naumann08-Patent,Yakorev11-NIMA,Naumann11-NIMA,Roeder12-JINST}, here the single-electron beam was produced by a scattering wire inserted in the beam line, allowing the measurements to run in parasitic mode, parallel to a different experiment requiring high beam intensity. It was periodically checked that the beam contained just one electron per bunch.

The electron beam is transmitted in air after leaving the evacuated beam line through a thin beryllium exit window. It then passes a 20$\times$20\,mm$^2$, 5\,mm thick plastic scintillator (read out on each side by PMTs called S1 and S2, respectively), then goes on to the scintillator to be studied (read out by SiPMs S7 and S8) and finally passes a 35$\times$25\,mm$^2$, 5\,mm thick plastic scintillator (read out by PMT S6). The sizes have been selected to be smaller than the scintillator under study, in order to enable the determination of the efficiency.

The electronics setup is the same as for the DAQ-based measurements with a picosecond laser (sec.\,\ref{sec:PiLas}), with the changes listed in the following. The SiPMs signals are again split by a linear fan-out unit: One branch goes to a CFD with 15\,mV threshold, the second to a CFD with 45\,mV threshold, and the third branch is delayed and passed to the QDC, with a charge integration time window of 250\,ns generated by the FPGA. For the PMT signals, a CFD threshold of 30-50\,mV is used. The DAQ trigger is given by S1$\wedge$S2$\wedge$RF. Here, RF denotes the radio frequency signal from the accelerator. The time reference is given by the RF signal, making the time resolution of S1, S2, and S6 irrelevant to the experiment. 

\begin{figure*}[bt]
\includegraphics[angle=0,width=\textwidth]{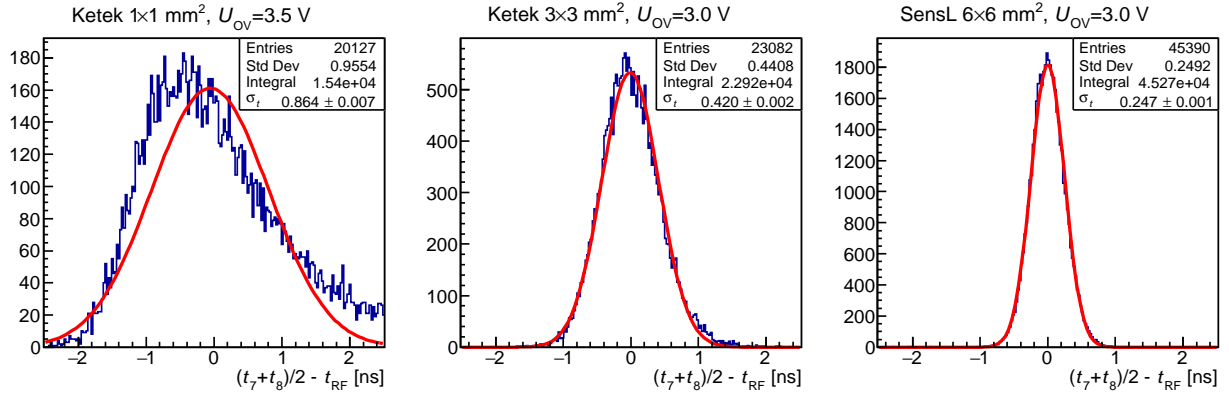}
\caption{\label{fig:Timeres_Threeruns} Average timing $(t_7+t_8)/2-t_{\rm RF}$ for the NeuLAND bar for three different sizes of the SiPM readout applied at each end. For better comparison, arbitrary offsets have been added to the time axes to center the peaks at zero. The red curves are Gaussian fits to the data, which have been used to determine the time resolution values $\sigma_t$ shown in the statistics boxes of the plots. See text for details.}
\end{figure*}
\begin{figure}[bt]
\includegraphics[angle=0,width=\columnwidth]{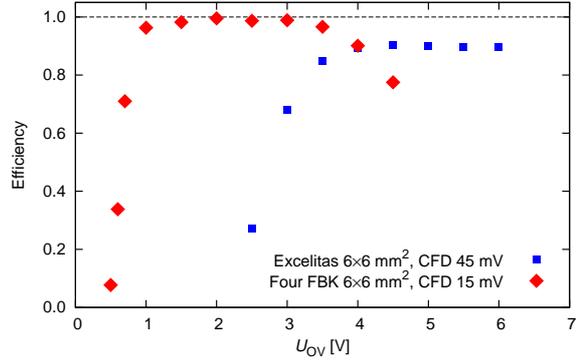}
\caption{\label{fig:Efficiency} Detection efficiency for an ELBE electron in a NeuLAND bar as a function of overvoltage for two examples studied.}
\end{figure}

For each combination of scintillator and SiPM studied, the scintillator was instrumented on both sides with a SiPM or a SiPM array of the same size, producer and series. The efficiency and time resolution were then studied as a function of the overvoltage $U_{\rm OV}$. For the final time resolution, the average time $(t_7-t_8)/2$ of right and left readout is used, in order to remove the time resolution contributed by the lateral size of the electron beam, $\sigma \approx$ 1\,cm. 

\subsection{Offline data analysis}

In the offline analysis, a valid electron beam event is assumed for events where S1, S2, and in addition also S6 show a valid signal in the timing channel. The additional requirement of a valid S6 time excludes electrons that are scattered to large angles in the plastic sample under study and also parasitic events caused by bremsstrahlung in different parts of the setup. 

The time-over-charge plot shows no significant time walk (i.e. no change of the mean time as a function of charge) for all the SiPMs studied here (fig.\,\ref{fig:Banana}). The separation between pedestal and first fired pixel is clearly visible, and second and third fired pixels can be separated, as well. Due to the intrinsic finite charge resolution of the device, this single-pixel capability is only evident in runs with a few fired pixels. For larger numbers of fired pixels, the charge spectrum is shifted to higher charges and smeared out, making individual pixels indistinguishable.

The detection efficiency is obtained by dividing the number of events in the main timing peak by the number of events in the reference scintillator S1/S2, for the same condition. The efficiency curve usually shows a plateau of 2\,V width (fig.\,\ref{fig:Efficiency}). The rising flank of the efficiency plateau is given by the onset of the avalanche process, when the breakdown voltage is exceeded and events exceeding the CFD threshold are generated. The falling flank is given by the dark rate, which rapidly rises with overvoltage, so high that the capability to detect incoming real photons is impeded. There is a strong voltage dependence both of the efficiency and of the dark rate outside the plateau. 

In the experiment, finally high-statistics data are taken at a point well inside the efficiency plateau and kept for further analysis (figs.\,\ref{fig:Banana} and \ref{fig:Timeres_Threeruns}). The time-over-charge plot shows only a very slight time walk (fig.\,\ref{fig:Banana}). For the small 1$\times$1\,mm$^2$ SiPM (fig.\,\ref{fig:Banana}), single firing pixels can even be distinguished on the charge axis. This is not the case for the larger SiPMs, where more pixels fire and thus the charge spectrum is smeared out. 

The average timing plot taken with the NeuLAND bar and the  1$\times$1\,mm$^2$ SiPM shows some tailing to late average times (fig.\,\ref{fig:Timeres_Threeruns}, left panel). The same effect is found in the Monte Carlo simulation (see following sec.\,\ref{sec:MonteCarlo}) for this combination of scintillator and SiPM. It might be due to a non-negligible probability of photons that are emitted at late times in the time emission profile of the scintillator material, but still are the first photon to be detected by the SiPM. This asymmetry gradually disappears for larger SiPM areas, and thus larger number of fired pixels (fig.\,\ref{fig:Timeres_Threeruns}, middle and right panels). 

\section{Monte Carlo simulations of light propagation and detection}
\label{sec:MonteCarlo}

In order to quantatively understand the dependence of the time resolution on the size of the SiPM used, simple Monte Carlo calculations have been performed. A simple customized code was developed for this purpose, taking advantage of the ROOT TRandom2 random generator. No full modeling of light propagation and photosensor is attempted here, as it has been done previously for smaller scintillator bars, where very good timing properties have been found \cite{Seifert12-PMB,Seifert12-IEEE}. Instead, the present approach is restricted to the main features of light propagation in a large scintillator. 

As a starting point, along the  track of the single ELBE electron scintillation photons are emitted in a random direction in three dimensions, using the collisional energy loss of 0.2\,MeV/mm in the plastic scintillator material for minimum ionizing particles and the datasheet value of 10,000 scintillation photons/MeVee. The time needed by the electrons to traverse the NeuLAND bar (170\,ps) is taken into account in the simulation. The starting time of each photon is determined by randomly sampling the assumed time distribution
\begin{equation}
\label{eq:Knoll}
\frac{N(t)}{N_{\rm max}} = \exp\left(-\frac{t}{\tau_{\rm decay}}\right) - \exp\left(-\frac{t}{\tau_{\rm rise}}\right)
\end{equation}
where $t$ is the time. For the rising and decaying flanks of the scintillation light, the respective datasheet values  $\tau_{\rm rise}$ = 0.9\,ns and $\tau_{\rm decay}$ = 2.1\,ns are used.

These scintillation photons are then propagated through the scintillator, assuming 99\% specular reflectivity at the material boundaries (a plausible guess \cite{Janacek12-IEEE}, as the precise choice of the reflector used for NeuLAND is only known to the manufacturer) and an attenuation length of 4000\,mm (the datasheet value). For photons that hit the SiPM, a random sampling with the photon detection efficiency (using a typical datasheet value of 35\%) is taken in order to determine which photons are detected. The time of detection of the individual photon is then given by the time of emission of this individual photon plus its travel time. This time is subsequently smeared out with a Gaussian function emulating a transit time spread in the electronics of $\sigma_{\rm TTS}$ = 50\,ps and then stored.

\begin{figure}[tb]
\includegraphics[angle=0,width=\columnwidth]{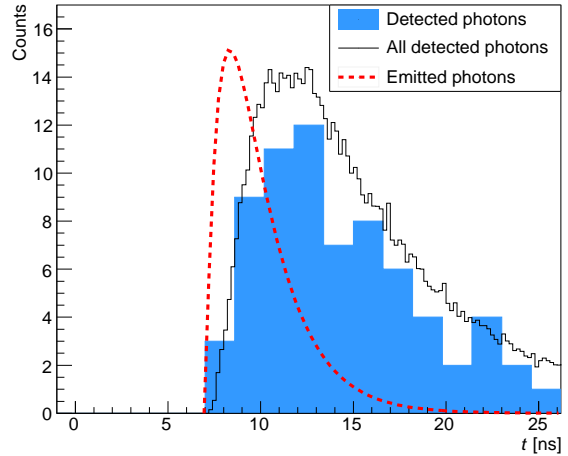}
\caption{\label{fig:Timing_MonteCarlo} Blue filled histogram: Monte Carlo simulated photon hits for a 30\,MeV electron traversing the NeuLAND scintillator, read out by a 3$\times$3\,mm$^2$ SiPM. Black empty histogram: All photon hits generated by many electron primaries (scaled down). Red curve: Initial time distribution at emission of the scintillation light.}
\end{figure}

For each electron primary, the detected photons are then filled into a time histogram (fig.\,\ref{fig:Timing_MonteCarlo}, blue filled histogram), and the rising flank time is determined by fitting with eq.\,(\ref{eq:Knoll}) and using a fraction of 0.22 times the maximum. As a byproduct, for each electron primary also the number of fired pixels ($n_{\rm FP}$ = 66 for the example shown in fig.\,\ref{fig:Timing_MonteCarlo}) is determined and stored. 

The histogram given by all scintillation photons detected, summed over all primary events, shows that the general time structure of the detected light is well described by eq.\,(\ref{eq:Knoll}), but with longer time constants. For the 2700$\times$50$\times$50\,mm$^3$ NeuLAND bar, $\tau_{\rm rise}$ = 2.0\,ns and $\tau_{\rm decay}$ = 6.2\,ns is found (fig.\,\ref{fig:Timing_MonteCarlo}, black empty histogram). For a smaller scintillator of 100$\times$42$\times$11\,mm$^3$, $\tau_{\rm rise}$ = 1.5\,ns and $\tau_{\rm decay}$ = 2.1\,ns are found instead. These effective left and right flank time constants are just a function of the scintillator size and shape and approximately independent of the area of the SiPM. For comparison, also the original time shape at emission is included in the plot ($\tau_{\rm rise}$ = 0.9\,ns and $\tau_{\rm decay}$ = 2.1\,ns; fig.\,\ref{fig:Timing_MonteCarlo}, red curve). 

\begin{figure}[tb]
\includegraphics[angle=0,width=\columnwidth]{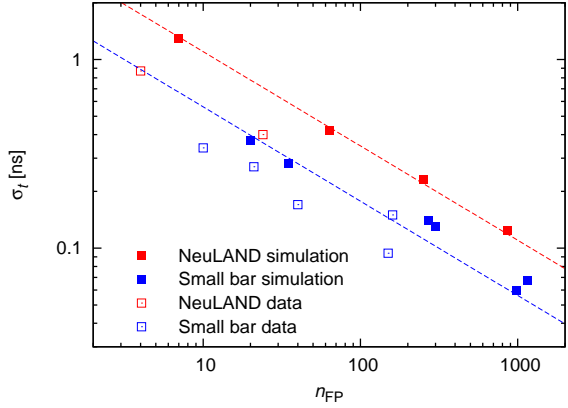}
\caption{\label{fig:Auffray} Time resolution $\sigma_t$ as a function of the number of fired pixels $n_{\rm FP}$, for the NeuLAND bar (red symbols) and the 100$\times$42$\times$11\,mm$^3$ bar (blue symbols). The simulation results (filled squares) and experimental data (open squares) from table~\ref{Table:Exp+Sim} are compared with the predictions using eq.~(\ref{eq:Auffray}) (red and blue dashed curves). See text for details.}
\end{figure}
\begin{table}[tb]
\resizebox{\columnwidth}{!}{%
\begin{tabular}{lrrrrr} \hline
Scintillator & SiPM & \multicolumn{2}{c}{Simulation} & \multicolumn{2}{c}{Experiment} \\
size in mm$^2$ & [mm$^3$] & $n_{\rm FP}^{\rm sim}$ & $\sigma_t^{\rm sim}$ & $n_{\rm FP}^{\rm exp}$ & $\sigma_t^{\rm exp}$ \\ \hline \hline
 100$\times$42$\times$11 & 1$\times$1 & 20 & 370 & 10 & 340 \\
 & 3$\times$3 & 270 & 140 & 40 & 170 \\
 & 6$\times$6 & 980 & 60 & 150 & 94 \\ \hline
 100$\times$11$\times$42 & 1$\times$1 & 35 & 280 & 21 & 270 \\
 & 3$\times$3 & 300 & 130 & 160 & 150 \\
 & 6$\times$6 & 1160 & 67 & $\ge$280 & 72 \\ \hline
 2700$\times$50$\times$50 & 1$\times$1 & 7 & 1300 & 4 & 870\\ 
 & 3$\times$3 & 64 & 420 & 24 & 400\\ 
 & 6$\times$6 & 250 & 230 & $\ge$125 & 240 \\
 & Four 6$\times$6 & 860 & 124 & & 135 \\ \hline \hline
\end{tabular}
}
\caption{\label{Table:Exp+Sim} Simulated and experimental values for the most probable number of fired pixels $n_{\rm FP}$ and the time resolution $\sigma_t$. All times are given in ps. See text for details.}
\end{table}

It is clear that the precision with which the rising flank of the distribution in fig.\,\ref{fig:Timing_MonteCarlo} can be determined is a function of the number of fired pixels, just as was the case for the laser-based measurements (sec.\,\ref{sec:PiLas} and fig.\,\ref{fig:PiLas_FiredPixels}). However, due to the above mentioned smearing out effect for both left and right flanks, for a given number of fired pixels the time resolution with a scintillator-based measurement is worse than when using just a laser. 

In order to quantify this effect, simulations based on the above considerations have been carried out for three scenarios with the 30\,MeV electron beam:
\begin{enumerate}
\item A small scintillator of 100$\times$42$\times$11\,mm$^3$ (100\,mm width between the two readout SiPMs $\times$ 42\,mm height perpendicular to the beam direction $\times$ 11\,mm length in the beam direction) EJ-200 material.
\item The same scintillator, but rotated by 90$^\circ$ so that the electron beam traverses 42\,mm length; it is denominated 100$\times$11$\times$42\,mm$^3$. 
\item The NeuLAND bar, for simplicity modeled as a 2700$\times$50$\times$50\,mm$^3$ cube of RP-408/EJ-200 material, thus neglecting the tapering at both ends.
\end{enumerate}
The results of these simulations are listed in table~\ref{Table:Exp+Sim} and compared to the electron beam data. For the data, runs at the efficiency plateau have been selected. It seems that the trends observed in the data for $n_{\rm FP}$ and $\sigma_t$ are well matched by the simulation. Also the absolute values for $\sigma_t$ match. 

However, the simulation consistently overpredicts the absolute number of fired pixels. This may be due to issues with the calibration of the experimentally determined number of fired pixels, which was done in the separate laser setup in a different room. Also, the starting time of the QDC integration time window might have been be somewhat different due to the different trigger conditions. Even though the scintillators under study were well polished, possible photon losses due to remaining surface roughness \cite{Gundacker13-JINST} are not included in the present model. 

Possible saturation effects \cite{Gruber14-NIMA} are expected to reduce the observed number of fired pixels to a fraction $f_{\rm Sat}$
\begin{equation}\label{eq:Gruber}
f_{\rm Sat} = \frac{n_{\rm Pixel}}{n^{\rm original}_{\rm FP}}\left[1-\exp\left(-\frac{n^{\rm original}_{\rm FP}}{n_{\rm Pixel}}\right)\right]
\end{equation}
 where $n^{\rm original}_{\rm FP}$ is the number of fired pixels without saturation effects. For the SiPMs studied here, the total number of pixels is $n_{\rm Pixel}$ = 576 (1$\times$1\,mm$^2$), 3600 (3$\times$3\,mm$^2$), 18980 (6$\times$6\,mm$^2$), and 75920 (array of four 6$\times$6\,mm$^2$ SensL), respectively. When taking the simulated number of fired pixels (Table~\ref{Table:Exp+Sim}) for $n^{\rm original}_{\rm FP}$, this leads to $f_{\rm Sat}$\,$\ge$\,0.96, i.e. less than 4\% correction due to saturation.

For the time resolution $\sigma_t$, again an inverse square root dependence is found both from the data and from the simulations. The simulated $\sigma_t$ values are well described by the following relation:
\begin{equation}\label{eq:Auffray}
\sigma_t = \sqrt{\frac{\tau_{\rm rise}\tau_{\rm decay}}{n_{\rm FP}}}
\end{equation}
A similar formula has been derived previously \cite{Auffray11-IEEE}. There, a very small scintillator was studied, where all photons reached the SiPM and could thus be detected with the SiPM photon detection efficiency. Assuming $\tau_{\rm rise}\ll\tau_{\rm decay}$, from the simple integration of the number of photons detected, the same relation as eq.~(\ref{eq:Auffray}) was found \cite{Auffray11-IEEE}, when considering that $n_{\rm FP}\equiv R/2$, where $R$ is the total number of produced photoelectrons. 

Here, the effective times $\tau_{\rm rise,fall}$ after propagation of the light to the end of the scintillator bars have to be used. Different from Ref.\,\cite{Auffray11-IEEE}, in the present work $\tau_{\rm rise,fall}$ have the same order of magnitude. Also, only a part of the scintillator surface is instrumented here, leading to unavoidable losses of light. Remarkably, the same relation still holds. It may allow to estimate the timing behaviour also of other fast scintillators, once the effective rise and decay times at the place of the photosensor, i.e. taking into account light propagation, are known. 

The above conclusions have been reached with a strongly simplified model, neglecting  afterpulsing and crosstalk \cite{Vinogradov15-NIMA}, which have to be taken into account for  counting rates that are higher than the present one. Also, it should be noted that contributions by diffuse light \cite{Gundacker13-JINST} have been neglected here.

\section{Electron beam measurements with a NeuLAND bar read out by SiPM arrays}
\label{sec:Arrays}

\begin{figure*}[btb]
\includegraphics[angle=0,width=\textwidth]{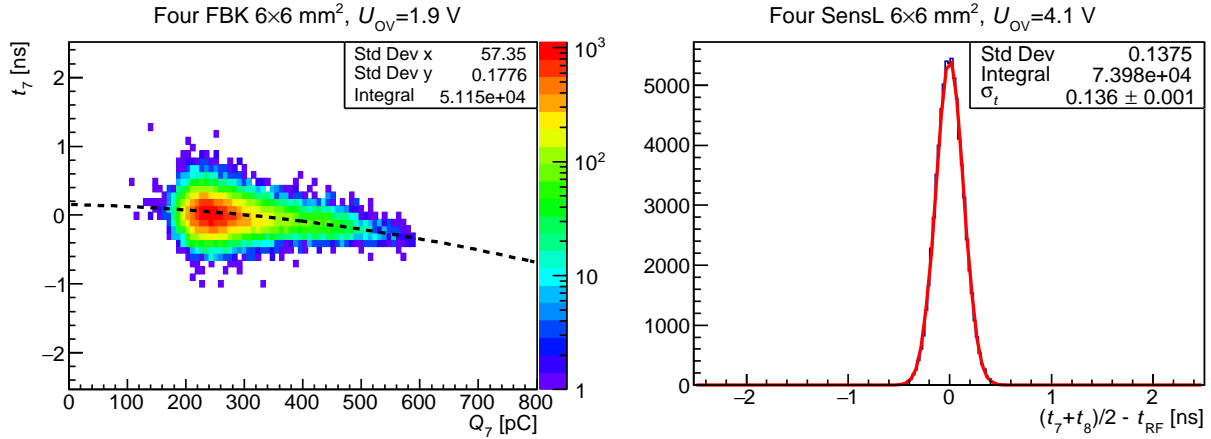}
\caption{\label{fig:Timing_1212} Data for two different arrays of four 6$\times$6\,mm$^2$ SiPMs, on the efficiency plateau. Left panel: Time-over-charge plot for one side of the FBK array; the dashed curve is the fitted time walk. Right panel: Averaged time resolution for the SensL array; the red curve shows a Gaussian fit used to determine the time resolution $\sigma_t$. See text for details.}
\end{figure*}

The time resolution trends for both the data, from different manufacturers, and the simulation, for a generic SiPM, indicate that the aimed NeuLAND time resolution of $\sigma_t<150$\,ps can be reached with a small array of four 6$\times$6\,mm$^2$ SiPMs on each end. In order to check this assumption, two couples of such arrays were made with SiPMs from different suppliers:
\begin{enumerate}
\item["S"] First, four SensL 6$\times$6\,mm$^2$ SiPMs (C series, 35\,$\mu$m pitch, SMT housing, $U_{\rm BD}\approx$ 25\,V) were included in a home-made array. Each of the individual SiPMs was separately supplied with bias voltage and had two outputs, a fast timing one and a slower spectroscopic output. The four timing outputs were simply added in an analogous summing circuit. The same was done for the four spectroscopic outputs.
\item["F"] Second, a specially made array of four FBK 6$\times$6\,mm$^2$ SiPMs (NUV type, 40\,$\mu$m pitch, PCB-FBK package, $U_{\rm BD}\approx$ 33\,V) was used. The outputs of the four SiPMs were separately provided and then simply added in an analogous summing circuit.
\end{enumerate}

The NeuLAND bar was then instrumented in a first experiment with two equal "S" (SensL) arrays, one on each end. In a second experiment, it was instrumented with two equal "F" (FBK) arrays, again one on each end. 

In each of the two experiments, the efficiency was  determined as a function of the overvoltage, and on the efficiency plateau ($\ge$95\% efficiency), long runs were taken for statistics (fig.\,\ref{fig:Timing_1212}). The experimental procedure was the same as described above in sec.\,\ref{sec:ELBE}.

It appears from the time-over-charge plot that there is again only a very small time walk, mainly in the high-charge tail of the distribution, which contributes only limited statistics (fig.\,\ref{fig:Timing_1212}, left panel). In order to check the effect of the walk on the determined time resolution, the time-over-charge plot was fitted with a second order polynomial for each side (dashed curve in fig.\,\ref{fig:Timing_1212}, left panel), and then the time for each side was corrected for its time walk before forming the average time. The final result changed by less than 4\,ps, so for the further analysis the time walk correction was omitted. 

The RF signal of the accelerator was found to contribute 26-30\,ps to the time resolution. When subtracting it quadratically from the experimental $\sigma_t$ value, the result decreases by 2-3\,ps, negligible for the present purposes. The time resolution data shown here are presented without this subtraction.

In order to check the behavior of $\sigma_t$ as a function of the SiPM area, individual SiPMs in the arrays were selectively switched off for some runs by reducing their bias below the breakdown voltage. Thus, arrays of one, two, and three 6$\times$6\,mm$^2$ modules were also studied. The data generally confirm the inverse square root dependence of $\sigma_t$ on the number of fired pixels, which is approximately proportional to the SiPM area for the present geometry.

The final efficiency and timing results, on the efficiency plateau, for the two sets of arrays with all four SiPMs active are:
\begin{itemize}
\item SensL C-series array: \\ (99$\pm$1)\% efficiency, $\sigma_t$ = 136$\pm$2\,ps. 
\item FBK NUV array: \\ (96$\pm$1)\% efficiency, $\sigma_t$ = 137$\pm$2\,ps.
\end{itemize}
The error bar for the time resolution results from the run-to-run reproducibility of the data. The statistical error bar from the Gaussian fit is always 1\,ps or below, which is negligible.

The two measured time resolution values for the arrays studied both fulfill the NeuLAND timing goal of $\sigma_t<$150\,ps. 

The present time NeuLAND resolution compares favorably to that obtained previously in $^{90}$Sr source experiments performed with a 100\,cm long monolithic RP-408 plastic scintillator, read out by 6$\times$6\,mm$^2$ SiPMs \cite{Kaplin15-PhysProc}. The experimental time resolution obtained in the 100\,mm long bar is comparable to the recently reported value for the PANDA-SciTil detector which is, however, planned to be implemented from many tiles of just 3\,cm size \cite{Brunner14-JINST}. 

\section{Summary and outlook}
\label{sec:Summary}

The readout of a large, monolithic scintillator bar with semiconductor-based photosensors called Silicon Photomultipliers has been studied experimentally and by simulations. 

As a first step, dedicated readout electronics was developed based on a two-stage amplifier providing a unipolar output signal that does not depend on the changing SiPM impedance. 

Subsequently, with picosecond laser-based measurements, it was shown that the assemblies made of SiPM and preamplifier show a competitive time resolution of $\sigma_t$ = 35\,ps if a sufficently high  number of light quanta is reached, so that 100 or more SiPM pixels fire. 

The setup was then further extended to a fast plastic scintillator bar read out on both ends by SiPM and preamplifier. The time resolution for 30\,MeV electrons, close to the minimum of ionization, was measured referred to the radio frequency signal of the ELBE superconducting electron accelerator, which is known to $\sigma_{\rm RF}$\,$<$\,30\,ps. The experiment was conducted using the one electron per bunch mode of ELBE. 

The ELBE data showed significantly worse resolution than the picosecond laser measurements. However, also for the electron beam data an inverse square root proportionality between the number of fired pixels and the time resolution was found to to hold. This observation was reproduced in a Monte Carlo simulation of light transport and detection. 

The simulation predicted a time resolution of $\sigma_t$\,$<$\,150\,ps for an array of four analog 6$\times$6\,mm$^2$ SiPMs reading out each end of a 2700$\times$50$\times$50\,mm$^3$ monolithic plastic scintillator. This prediction was subsequently verified in an ELBE electron beam experiment, for two different SiPM arrays from two manufacturers. The final time resolution reached was $\sigma_t$\,=\,137\,ps for 30\,MeV electrons. This value is well within the aimed time resolution for the NeuLAND neutron detector at FAIR \cite{NeuLAND-TDR11}.

It is interesting to note that the area instrumented on each side is just 144\,mm$^2$. This is just 6\% of the total NeuLAND area of 50$\times$50\,mm$^2$, and 30\% of the tapered end area of NeuLAND used for readout. 

This finding opens the road for a possible re-instrumentation of NeuLAND with small SiPM arrays, instead of the presently adopted \cite{NeuLAND-TDR11} fast timing photomultipliers. That would allow to make use of the advantages of SiPMs, such as much lower bias voltage requirement and insensitivity to magnetic fields, in a radioactive ion beam experiment. 

\section*{Acknowledgments}
The authors are indebted to Claudio Piemonte and Fabio Acerbi (Fondazione Bruno Kessler, Trento, Italy) for generously supplying the arrays of four FBK-NUV 6$\times$6\,mm$^2$ SiPMs, to Andreas Hartmann (HZDR) for technical support, and to Konstanze Boretzky (GSI Darmstadt, Germany) for helpful comments. The authors thank all their colleagues from the NEDENSAA network for stimulating discussions.  --- Financial support by NupNET NEDENSAA (BMBF 05 P 09 CRFN5), GSI F\&E (DR-ZUBE), the Helmholtz Detector Technology and Systems   Platform (DTS), and the European Union (MUSE network, contract no. 690835) is gratefully acknowledged. S.R. is an associate member of the Helmholtz graduate school HGS-HIRe for FAIR. 

\section*{Dedication}
This work is dedicated to the memory of our late colleague Mathias Kempe.

\end{document}